\documentclass[final,3p,times,twocolumn]{elsarticle-mod}

\usepackage{amssymb}

\usepackage[hyphens]{url}

\usepackage{amsmath}

\usepackage{color}
\usepackage{xtab}
\usepackage{overpic}
\usepackage[para]{threeparttable}
\usepackage{etoolbox}
\usepackage{pict2e}

\usepackage{fancyhdr}

\newcommand{\MARKI}[1]{#1}
\newcommand{\MARKII}[1]{#1}
\newcommand{\MARKIII}[1]{#1}

\appto\TPTnoteSettings{\footnotesize}

\journal{NSS Space Settlement Journal}

\begin{document}

\begin{frontmatter}



\title{Shielded Dumbbell L5 Settlement}


\author[FMI,Tartu,Aurora]{Pekka Janhunen\corref{cor1}}
\ead{pekka.janhunen@fmi.fi}
\ead[url]{http://www.electric-sailing.fi}

\address[FMI]{Finnish Meteorological Institute, Helsinki, Finland}
\address[Tartu]{Also at Tartu Observatory, Tartu, Estonia}
\address[Aurora]{Also at Aurora Propulsion Technologies Oy, Espoo, Finland}
\cortext[cor1]{Corresponding author}

\begin{abstract}
  We present a two-sphere dumbbell configuration of a rotating
  settlement at Earth-Moon L5. The
  two-sphere configuration is chosen to minimize the radiation
  shielding mass which dominates the mass budget. The settlement has
  \MARKII{max} 20 mSv/year radiation conditions and 1 g artificial gravity. If made for 200 people, it
  weighs 89000 tonnes and provides 60 m$^2$
  of floorspace per person. The radiation shield is made of asteroid
  rock, augmented by a water layer with 2\,\% of the mass for neutron
  moderation, and a thin
  boron-10 layer for capturing the thermalized neutrons.
  We analyze the propulsion options for moving the material from asteroids to L5.
  The FFC Cambridge process can be used
  to extract oxygen from asteroid regolith. The oxygen is then used as
  Electric Propulsion propellant. One can also find a
  water-bearing asteroid and use water for the same purpose. If one wants to avoid propellant extraction, one can use
  a fleet of electric sails. The settlers fund their
  project by producing and selling new settlements by zero-delay teleoperation
  in the nearby robotic factory which they own.  The economic case
  looks promising if LEO launch costs drop below $\sim$ \$300/kg.
\end{abstract}

\begin{keyword}
L5 space settlement \sep
radiation shielding


\end{keyword}

\end{frontmatter}





\section{Introduction}

To live permanently in space, a human being needs air, food, radiation
shielding, earthlike gravity, and a sufficient number of fellow
settlers and living space. All requirements can be satisfied in
rotating free-space habitats made of asteroid or lunar materials
as proposed by Gerard O'Neill in his pioneering works \citep{ONeill1974,ONeill1977}. Such unplanetary living is attractive
because it offers a way to avoid common natural hazards such as
hurricanes, volcanism, earthquakes and wildfires. It is also
attractive from the longer term population and economic growth points
of view. There is so much material in the asteroid belt
($2.4\cdot 10^{21}$ kg) that if made into settlements, it allows
several orders of magnitude growth in the human population.

One could build an orbital settlement in equatorial low Earth orbit
(ELEO) with much lower mass than elsewhere, because in ELEO the
Earth's magnetic field protects rather well against cosmic rays and
solar protons \citep{GlobusEtAl2017}. However, in LEO there is the
risk of orbital debris. For example, recently the insurance company
Assure Space stopped offering collision risk coverage policies in LEO
\citep{SpaceNewsAssureSpace}. There is also the issue of having to
perform a targeted reentry when done with the facility. Letting it
fall freely would be a public safety issue for people living near the
equator.


To avoid these issues, in this paper we consider a settlement at the
Earth-Moon Lagrange L5 (or L4) point. The L5 point offers Apollo-like
short traveltime from Earth, so that the transfer vehicle does not
need much radiation shielding. \footnote{We write ``L5 point'' for
  brevity. In reality the orbit wanders around the L5 or L4 point with
  large amplitude \citep{HouEtAl2015}.} Satellite orbits around $\sim$
8--10 Earth radii are also an alternative, but they have
somewhat larger delta-v for bringing asteroid material than
L5. The Moon makes most orbits in $\sim$ 15--150 Earth radii
unstable in the long run. L5 is an exception. Distant
translunar orbits would also be possible, but they exhibit longer
traveltimes from Earth. Longer manned trips increase the mass of the
transfer vehicle through the radiation shielding and the life support system.

The structure of the paper is as follows. First we consider radiation
shielding. Then we establish that because radiation shielding
dominates the mass, an optimal entry-level configuration is a dumbbell
comprising two spheres. We point out that only the radiation shielding
mass needs to be sourced from asteroids in the first phase. We analyze
a number of propulsion options for moving the material and do costing
analysis. We close the paper by discussion and summary.

\rfoot{\textit{NSS Space Settlement Journal}}
\pagestyle{fancy}
\renewcommand{\headrulewidth}{0pt}
\rhead{}
\lhead{}

\section{Radiation shielding}

Galactic cosmic rays (GCRs) have higher energies than solar energetic
particles. For good radiation protection, the GCR flux must be
significantly suppressed, and this requires several tonnes of mass per
square meter. At such shielding thicknesses the solar energetic
particles are suppressed almost entirely so they can be ignored in the
analysis.

\citet{GlobusAndStrout2017} used the OLTARIS tool to simulate the GCR
equivalent radiation doses (millisievert rates) behind various
thicknesses of different materials (\citet{GlobusAndStrout2017}, Table
2). They recommend 20 mSv/year as the equivalent dose level during
the solar minimum (of 2010) when the galactic radiation is at maximum, a value which
we also adopt here.

Here we also use the OLTARIS tool. The tool supports two geometries: a
slab and a sphere. In the slab geometry, the dose between two adjacent
infinite plates is predicted, where each plate has the given shielding
thickness. In the sphere geometry, the program predicts the dose at
the center of a solid sphere whose radius is equal to the given
shielding thickness. The dose predicted using the slab geometry is
smaller (typically about two times smaller) than that using the sphere
geometry, because in the slab geometry, many of the cosmic rays enter
obliquely to the shield and so move a longer distance within the
shield. For a hollow sphere, in the limit where the inner radius is
much larger than the shell thickness, the dose at the inner wall
approaches asymptotically the slab geometry prediction. The dose at
the centerpoint of a hollow sphere does not depend on the sphere's
inner radius, so it can be calculated by assuming zero inner
radius. Thus for a hollow sphere, the dose at the center is larger
than the dose near the walls. \MARKIII{The reason is that center-reaching cosmic
  rays pass through the shell perpendicularly regardless of their arrival direction,
  whereas near the wall typical
  rays must pass through the shell obliquely, experiencing more attenuation.}

For radiation shields, Z-grading is in general useful, that is, using
high-Z materials as the outer layer and progressively lower Z
materials as one goes inwards. We want the bulk of the shield to be
asteroid regolith. We also need structural material, which we assume
to be steel, which is iron to a good approximation. We put the steel
as the outermost layer since iron's mean atomic mass is larger than
that of regolith. Inside the regolith we put a layer of water whose
mass is 2\,\% of the mass of the regolith and water combined. It is
the intention that this amount of water can be obtained from the
asteroid regolith by heating it. If not, the water can be brought from
Earth. The dry regolith layer is modeled by OLTARIS' lunar regolith
option.

Part of the equivalent dose consists of neutrons spallated from the
regolith by cosmic rays. The water layer moderates these neutrons. As
the innermost layer we add a thin 1 kg/m$^2$ layer of boron-10. (If
one wants to avoid isotope separation, one can use 5 kg/m$^2$ of
natural boron which is 20\,\% B-10 and 80\,\% B-11.) This
isotope absorbs neutrons efficiently, especially thermal neutrons. The
water moderates the neutrons to be absorbed by the boron.

Table \ref{tab:wall} defines the layers of our wall. This shield was
designed to \MARKII{limit radiation to} 20 mSv/year equivalent dose at the center of the
sphere during worst case, i.e., solar minimum. The equivalent doses were
computed for ``Female Adult Voxel'' phantom \citep{FAX}.

\begin{table}[htb]
\caption{20 mSv/year shield using 2\,\% of water.}
\begin{tabular}{lll}
\hline
Material, \MARKI{thickness} & kg/m$^2$ & Role \\
\hline
Iron, \MARKI{2.3 cm} & 180 & Structural wall \\
Dry regolith, \MARKI{3.34 m} & 8683 & Shield \\
Water, \MARKI{17.7 cm} & 177 & Neutron moderator \\
Boron-10, \MARKI{0.4 mm} & 1 & Neutron absorber \\
\hline
Total, \MARKI{3.5 m} & 9041 & \\
\hline
\end{tabular}
\label{tab:wall}
\end{table}

For the solar maximum (of 2001) conditions, the equivalent dose is
25\,\% smaller (14.87 mSv/year). For solar minimum, a slab geometry
calculation shows that near the inner wall the equivalent dose is
50\,\% smaller (9.97 mSv/year) than at the center. During solar
maximum the inner wall equivalent dose drops to 7.53 mSv/year.

\citet{GlobusAndStrout2017} also require that the absorbed dose for
pregnant women be less than 6.6 mGy/year, or 5 mGy per pregnancy. In
our case this condition is satisfied since the absorbed dose at the
center of the sphere during solar minimum is 4.0 mGy/year.

\section{Mass-optimal geometry}

\begin{figure}[htb]
\centering
\begin{overpic}[width=0.49\textwidth]{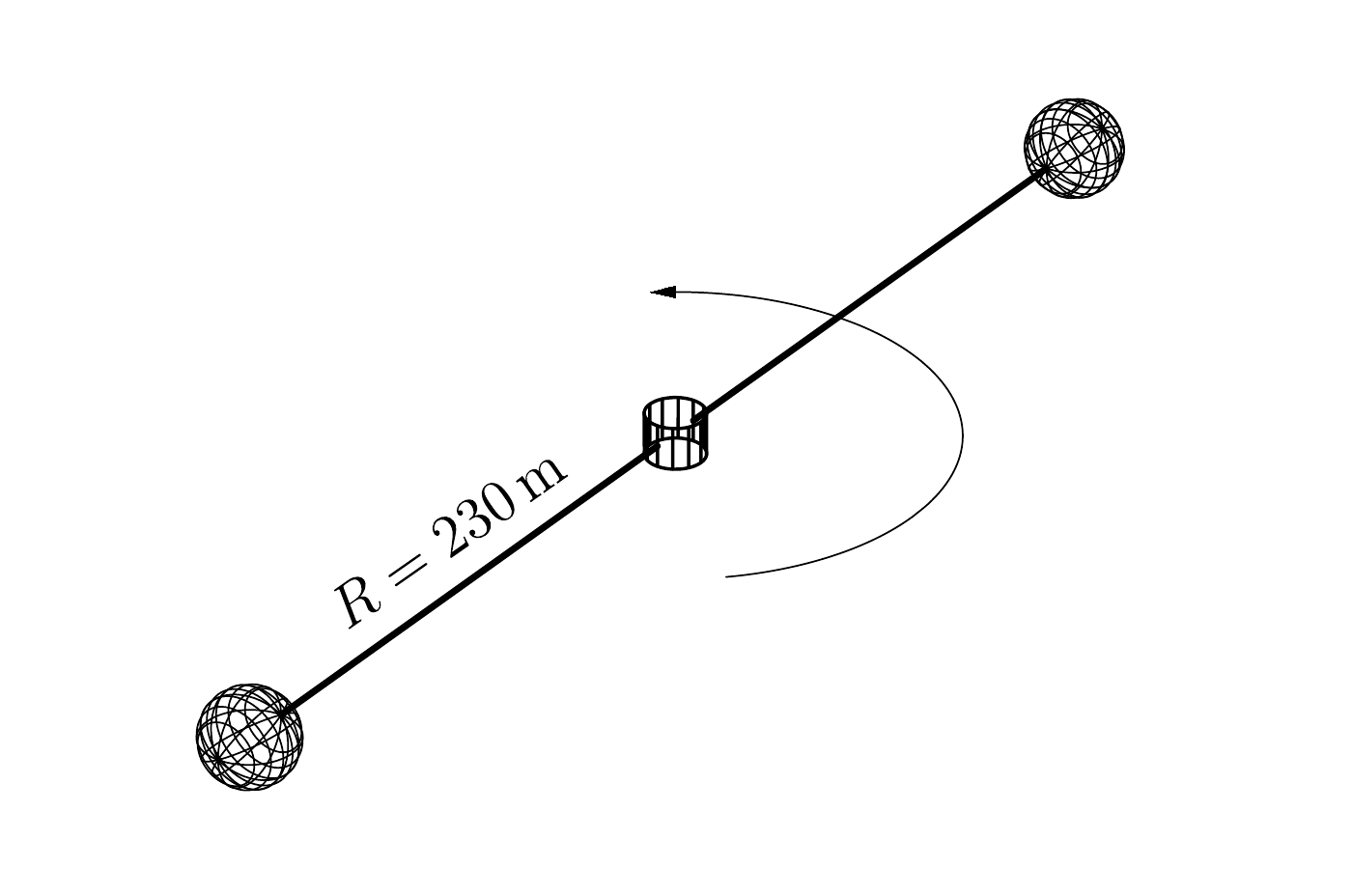}\put(3,62){a)}\end{overpic}
\begin{overpic}[width=0.49\textwidth]{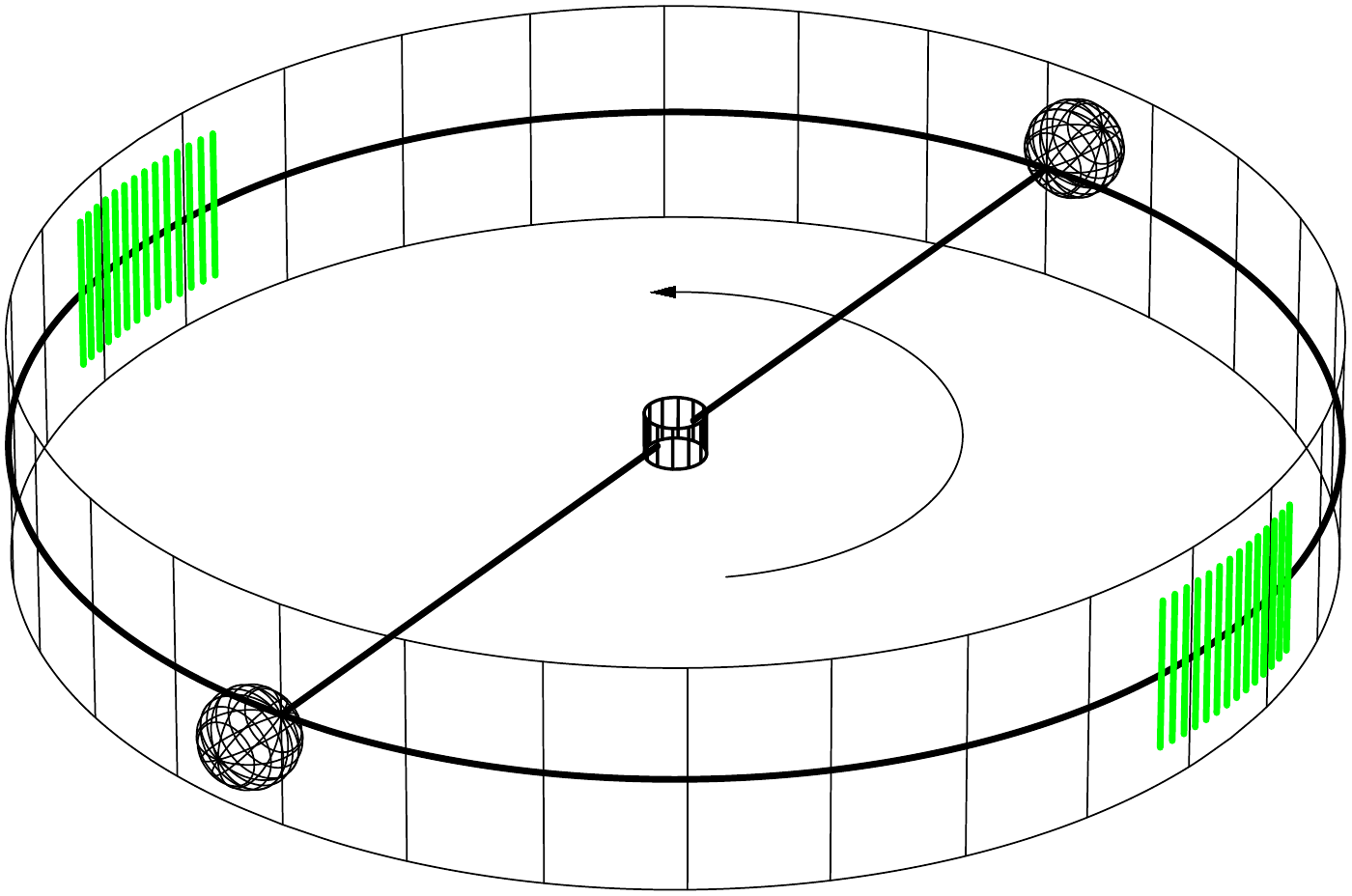}\put(3,62){b)}\end{overpic}
\caption{(a) Dumbbell settlement with elevator shaft and central
  docking ports, (b) with ringroad, cylindrical solar panels
  and greenhouse areas indicated (green).}
\label{fig:Settlement}
\end{figure}

A sphere has the minimal surface area per volume and is thus the best
geometry for minimizing shielding mass.  To include artificial
gravity, we need two spheres rotating about each other in a dumbbell
configuration (Fig.~\ref{fig:Settlement}a). We select a baseline
rotation rate of 2 rpm (revolutions per minute), which is probably a
conservative choice regarding avoidance of motion sickness
\citep{GlobusAndHall2017}. With 1 g artificial gravity, 2 rpm
corresponds to rotation radius of $R=230$ m.  The main structural
element is the truss that connects the spheres. It carries the
centrifugal load of the heavy spheres. It also acts as the shaft of an
elevator that provides access from the living spheres to a central
docking port. There are two docking ports for redundancy, upward and
downward in Fig.~\ref{fig:Settlement}.  The docking ports are
essential because they are the way to enter and exit the
settlement. To ease the docking of the connecting spacecraft, the docking
ports are located on the axis of rotation.

To move from one sphere to the other, one can use the elevators, but
then one experiences temporary weightlessness in the central region,
which is an inconvenience. To eliminate this problem and to
provide redundancy in routing, we add a pressurized ring-shaped tube
that hosts a road. The ringroad is at constant radial distance from
the rotation axis so its user does not 
\MARKIII{need to move uphill or downhill in artificial gravity}.

To produce food as part of the closed ecosystem, we add
artificially illuminated greenhouses (Fig.~\ref{fig:Settlement}). The greenhouses rest on the
cylinder on which solar panels are mounted. The greenhouses are served
by the ringroad and we place them 90$^{\circ}$ off the heavy living spheres
to make the azimuthal mass distribution more uniform. We assume that food production needs \MARKII{16}
kW of electrical energy per person, so \MARKII{3.2} MW total for population of
200. This corresponds to \MARKII{2500 kcal per day per person of food plus 30\,\%
margin, and $1$\,\% efficiency in converting greenhouse electrical energy} into
edible energy of the crops. Most of the
energy is dissipated inside the greenhouses and radiated into space
from their roofs. To keep the heat transfer passive and thus reliable, the radiator
must be cooler than the greenhouse. At radiator
temperature of +4 C, the emitted thermal power per area is 300 W/m$^2$ at
emissivity of 0.9. Thus to dissipate \MARKII{3.2} MW of power one needs \MARKII{10560}
m$^2$ of greenhouse roofs. The greenhouses are stacked
in as many layers as is needed to yield the wanted amount of
radiated cooling power per roof area. The green areas in Fig.~\ref{fig:Settlement}b
show the greenhoused areas. \MARKII{Pressure containment of the
  greenhouses also contributes to structural mass. The mass is
  proportional to greenhouse total volume. To calculate the volume, we
  assume 50 W/m$^3$ of volumetric power dissipation and 1 bar
  greenhouse pressure.}

Agriculture is \MARKII{performed} robotically because the greenhouses are
outside of the thick radiation shields.  \MARKIII{We make an
  assumption that agriculture
  works despite GCR.
  To what extent this is true depends on many factors, one
  of which is plant lifetime. Plants that grow rapidly from seeds are
  less likely to have problems due to radiation-induced mutations than long-lived trees, for
  example. More research is needed on this point.}

We place the greenhouses so that
they can be served by the ringroads. The ringroads are used not only by
people, but also by wheeled robots that move crops to the living
spheres and human wastes back to the greenhouses. The indicated area
of the cylindrical solar panel in Fig.~\ref{fig:Settlement}b is larger
\MARKII{(by factor of 2)} than what is needed to produce \MARKII{3.2} MW, because the settlement \MARKII{needs}
power also for other purposes than food production. \MARKIII{The
  spin axis is perpendicular to the ecliptic plan so that the solar
  panels are all the time optimally illuminated.} The power system
\MARKII{is} a \MARKII{rather small} fraction of the total mass, which is dominated by
radiation shielding mass (97\,\%) and structural mass
(\MARKII{2.6}\,\%).

\begin{figure*}[htbp]
\centering
\includegraphics[width=0.8\textwidth]{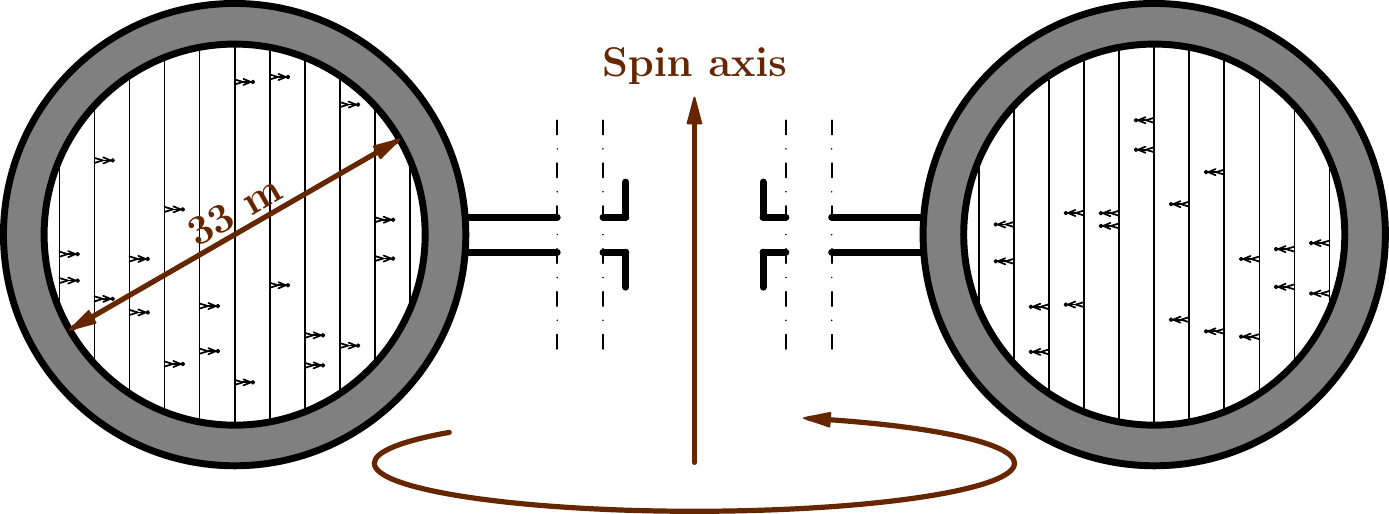}
\caption{Cross-sectional view.}
\label{fig:Schematic}
\end{figure*}

Figure \ref{fig:Schematic} shows a cross-sectional view of the
spheres.

The inhabitants live in the centrifugally produced artificial
gravity, with their heads towards the axis of rotation. Each sphere has inner diameter
of 33 m. \MARKII{In this section we give the baseline values of
  the parameters and motivate them later.} With effective room height of 3 m it contains ten
floor levels and provides 6000 m$^2$ of floorspace area for its 100
inhabitants so that each person has 60 m$^2$ of living area. For
example, it can be 25 m$^2$ of private area per person, 32 m$^2$ of public
and working areas including corridors, and 3 m$^2$ (5\,\%) taken by walls.
Tables \ref{tab:params200} and \ref{tab:budget200} list the main parameters
of the 200 person settlement.

\begin{table}[htb]
\centering
\caption{Parameters of 200 person settlement.}
\begin{tabular}{ll}
\hline
Population & 200 \\
Floorspace per person & 60 m$^2$ \\
Sphere inner diameter & 33 m \\
Rotation radius & 230 m \\
Steel tension & 800 MPa (30\,\% of limit) \\
Radshield mass per area & 9.04145 t/m$^2$ \\
Radshield density & 2.6 g/cm$^3$ \\
Radshield thickness & 3.5 m \\
\hline
\end{tabular}
\label{tab:params200}
\end{table}

For relatively small settlements such as 200 inhabitants, the density
of the regolith affects the mass: the denser the shield, the less of
it is needed because the sphere's radius is not that much larger than
the shield thickness. Bulk densities of asteroids vary in rather wide
range. Asteroid Eros is stony and with bulk density of 2.67 g/cm$^3$
\citep{YeomansEtAl2000}. Smaller stony asteroids are less compressed
by gravity and they can have lower densities; for example Itokawa has
1.95 g/cm$^3$. The minerals themselves would allow even higher
densities. The most abundant minerals are SiO$_2$ whose density is
2.65 g/cm$^3$ and MgO which is 3.6 g/cm$^3$. Iron-rich minerals are
often even denser. We use the value 2.6 g/cm$^3$ which is a bit less
than Eros' bulk density. Reaching this density requires the fragmented
rock to be compressed by vibrating, pressing or melting.

\begin{table}[htb]
\centering
\caption{Power and mass budgets of 200 person settlement.}
\begin{tabular}{llll}
& Total & Fraction & Per person \\
\hline
Power & \MARKII{3.2} MW & & \MARKII{16} kW \\
\hline
Radshield & 86423 t & \MARKII{96.9}\,\% & \MARKII{432} t \\
Structural & \MARKII{2270} t & \MARKII{2.55}\,\% & \MARKII{11.35} t \\
Other & \MARKII{493} t & \MARKII{0.55}\,\% & \MARKII{2.47} t \\
\hline
Total & \MARKII{89186} t & 100\,\% & \MARKII{446.0} t \\
\hline
\end{tabular}
\label{tab:budget200}
\end{table}

The mass efficiency of radiation shielding increases if the population
is increased, because the spheres become larger. When making the
spheres larger, however, the difference in artificial gravity between
the top and the bottom increases, unless one also increases the
rotation radius. When scaling up we require that the maximum gravity
is not more than 10\,\% larger than the average, and the minimum is
not more than 10\,\% smaller.

Increasing the rotation radius increases the structural mass fraction,
because the sphere-supporting trusses become longer. We propose that
the structural mass is piano wire steel. This material is 99\,\% of
iron, and iron is abundant on asteroids. In the first phase the
structural material is brought from Earth, but in later stages it can
be sourced from asteroids. We also set a requirement that the
structural fraction does not increase beyond 17\,\%, because the majority of
stony asteroids have more iron than this limit. Nothing prevents an
even larger iron fraction, but the only drawback is that then one may have to start abandoning some of the
asteroid material as waste. When the rotation
radius is increased to 1.6 km, the 17\,\% limit is
reached. If one wants to further increase the population without
increasing the max/min gravity difference beyond $\pm 10$\,\%, one can
add more spheres. Configurations of up to $\sim 30$ spheres are still
more mass efficient than an uninterrupted torus.

\begin{figure}[htbp]
\centering
\includegraphics[width=0.99\columnwidth]{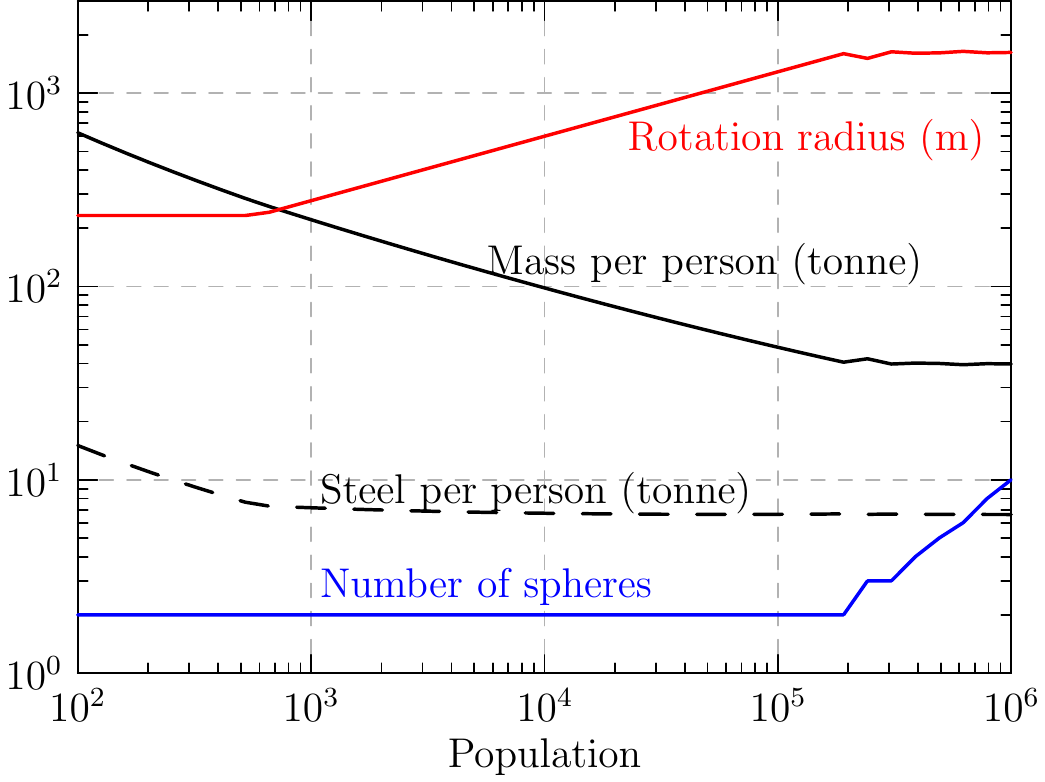}
\caption{Scaling as function of population.}
\label{fig:Logplot}
\end{figure}

Figure \ref{fig:Logplot} shows the rotation radius, mass per person,
structural mass (steel) per person and the number of spheres as a
function of population. For each population, the mass-optimal
configuration was found automatically. The rotation radius is the
constant 230 m up to 600 people, after which it grows in order to
avoid more than $\pm 10$\,\% difference in gravity between sphere top
and bottom. The mass per person is inversely proportional to sphere
radius, so that it is proportional to power $-1/3$ of the sphere
volume and population. For small spheres, the dependence is slightly
steeper because the radiation shield thickness is not negligible in
comparison to sphere radius in this regime. For most of the population
range, the structural mass per person is almost constant. This is
because two competing effects nearly cancel each other. A larger
rotational radius makes the truss longer, but it also enables larger
spheres which yields less radiation shielding mass per person and
therefore smaller carried load for the truss, per person. For
population less than 600, the rotation radius is constant because we
do not allow faster than 2 rpm rotation rate. 

With the employed parameters, two is the optimal number of spheres up
to population of $2\cdot 10^5$.

Table \ref{tab:requirements} summarizes the employed requirements.

\begin{table}[htb]
\centering
\caption{Employed limitations.}
\begin{tabular}{lll}
Parameter & Value & Motivation \\
\hline
Rotation rate & $\le 2$ rpm & No dizziness \\
Difference in gravity & $\le \pm 10$\,\% & Life convenience \\
Structural fraction & $\le 17$\,\% & No wasted \\
& & asteroid material \\
\hline
\end{tabular}
\label{tab:requirements}
\end{table}

\section{Material sourcing}

Most of the material is asteroid rock or regolith used for radiation
shielding. This fraction is 97\,\% in the baseline case of 200
people. Thus one can already reach asteroid mass ratio of 30
\MARKII{[$=97/(100-97)$]} by bringing only rock from the asteroids.

Most of the rest is structural material and most of it is used for the
two trusses that support the living spheres. Tensile strength is the
most important mechanical property required; compressive strength is
less important.  We recommend to use piano wire steel, which is 99\,\%
iron.  The tensile strength is 2.6-3 GPa and the production process
stems from the 19th century so it is not high-tech. 
If the structural material is sourced from asteroids, the mass ratio
increases to 170 in the baseline case of 200 people (Tables
\ref{tab:params200} and \ref{tab:budget200}). Sourcing structural
material from asteroids requires strict quality control
because the structural parts are life-critical.

To increase the mass ratio further, one could import water and carbon
from asteroids as raw materials for growing the biomass that
circulates between plants and people.

\section{Propulsive transfer of materials}
\label{sect:propulsion}

Most of the mass (97\,\% for a 200-person dwelling) is radiation
shielding, for which there are no structural or other requirements so
that it can be any unprocessed or processed asteroid rock. Thus the
main challenge is propulsion: how to transfer material from asteroids
or the Moon to the L5 orbit. Here we briefly analyze some of the potential
methods.

\subsection{Lunar material}

Lunar material could be lifted by an electromagnetic mass driver as
envisioned by O'Neill \citep{ONeillAndKolm1980}. The mass driver would
be a large investment. The electrical energy of the shot must be
stored in large capacitor banks, which is a major cost item. The fixed
shooting direction tends to reduce the flexibility regarding target
orbit. The centralized nature of the facility is a potential
reliability concern.

Lunar material could also be lifted by a sling
\citep{BakerAndZubrin1990,Landis2005}, which would be much lighter
infrastructure than the electromagnetic mass driver and it avoids the
use of capacitor banks. However, because the Moon rotates while the
plane of the rotating sling stays inertially fixed, the sling crashes
to the surface after some time, unless prevented by propulsion or
other means. The time to crashing depends on the parameters, but is
typically inconveniently short.

To lift lunar material, it is often proposed to make LH$_2$/LOX
propellant from the water ice that exists in the polar lunar
regions. However, the estimated H$_2$O resource is only a few times
$10^{11}$ kg \citep{EkeEtAl2009}. This amount is not sufficient for
long-term use. For example, a settlement with $10^6$ people
corresponds to mass of $3.5\cdot 10^4$ kg per person (Fig.~\ref{fig:Logplot}),
i.e.~total mass of $3.5\cdot 10^{10}$ kg, which is already $\sim
10$\,\% of the total lunar water resource\footnote{The chemical propellant mass
  is of the same order of magnitude as the payload mass in the lunar
  case.}.

\subsection{Asteroid material transferred by O$_2$ electric propulsion}
\label{subsect:O2EP}

As shown recently \citep{LomaxEtAl2020}, the so-called FFC Cambridge
electrolytic process can be used to separate earthly, lunar or
asteroid rock into oxygen gas and a solid residue comprising metals
and silicon. The oxygen can be used as Electric Propulsion propellant
\citep{AndreussiEtAl2019}. The O$_2$ Electric Propulsion technology is
currently under development in the context of Air-Breathing Electric
Propulsion \citep{AndreussiEtAl2019}, which is enabling technology for
very low orbiting satellites that are naturally immune to orbital
debris and do not \MARKII{generate} new debris.

The FFC Cambridge process requires calcium chloride electrolyte. The
electrolyte can be \MARKIII{recycled}, but the initial amount must be brought
from Earth. Chlorine is a rather rare element on asteroids. In the
proof of concept experiment of \citet{LomaxEtAl2020}, 1.6 kg of
CaCl$_2$ was used to process 30 grams of lunar regolith
simulant. According to the newest results \citep{LomaxEtAl2019},
$\sim 75$\,\% of the total oxygen was extracted after 16 hours in the
reactor. Thus, during one year, 545 batches can be processed,
altogether processing 16.4 kg of regolith and liberating 4.3 kg of
O$_2$, if the total oxygen content of the rock is 35\,\%. Thus for one
year, the mass ratio (O$_2$ : CaCl$_2$) is (4.3 : 1.6) = (2.7 :
1). Because the process demonstration was intended only as a proof of
concept and was not optimized, it is likely that the amount of
electrolyte and/or the throughput time can be improved, maybe
significantly.

\subsubsection{\MARKII{Delta-v from asteroids}}

To estimate the delta-v from a given asteroid to L5, we compute the
optimal Hohmann transfer delta-v from the asteroid to circular zero
inclination 1 au heliocentric orbit, consisting of two or three
impulsive burns, whichever strategy gives the smallest delta-v. The
burns set the aphelion, the perihelion and the inclination. In
reality, since we are considering low-thrust Electric Propulsion, the
burns are not impulsive and therefore they are not optimal. On the
other hand, in reality one could make use of lunar flyby maneuver to
kill up to $1.6$ km/s of of the incoming hyperbolic excess speed.
Because the effects work in opposite directions regarding the needed
delta-v, we think that the impulsive Hohmann transfer delta-v gives a
useful approximate measure of the low-thrust delta-v.

\begin{figure}[htb]
\includegraphics[width=0.99\columnwidth]{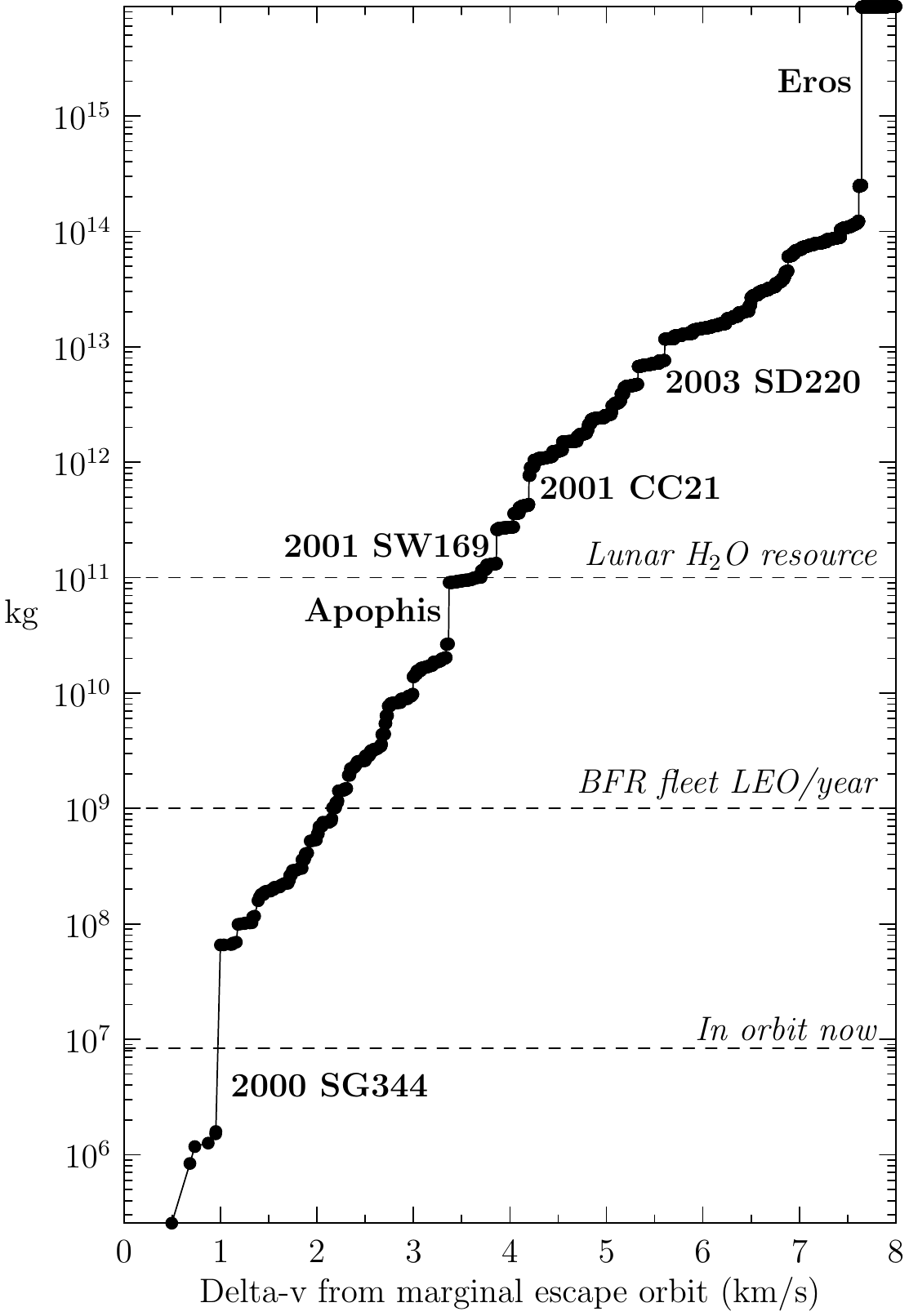}
\caption{
Cumulative mass of known asteroids as a function of delta-v.
}
\label{fig:Cum}
\end{figure}

Figure \ref{fig:Cum} shows the cumulative mass in known asteroids
sorted by the delta-v computed as just explained. The masses in
Fig.~\ref{fig:Cum} are based on the tabulated absolute magnitudes in
JPL Small Body Database by assuming albedo of 0.15, density of 2
g/cm$^3$, and spherical shape. The cumulative mass jumps at certain
large and well accessible asteroids, some of which are marked in
Fig.~\ref{fig:Cum}. To build the first settlement, we need 78000
tonnes of asteroid rock, which is only little larger than the
estimated mass of asteroid 2000 SG 344, which in one source has been
estimated as $7.1 \times 10^7$ kg \citep{NASAwebsite}. To be
conservative, however, we shall assume the use of asteroid Apophis
whose mass is 3 orders of magnitude larger. Apophis is also one of the
potentially hazardous asteroids, so reducing its mass is not harmful
from the planetary defense perspective.

The delta-v from Apophis is 3.37 km/s. For the transfer spacecraft
using O$_2$ electric propulsion, we assume the parameters listed in
Table \ref{tab:transferSC}.

\begin{table}[htb]
\centering
\caption{Parameters of the electric propulsion transfer spacecraft.}
\begin{tabular}{ll}
\hline
Delta-v          & 3.37 km/s \\
Specific impulse & 2500 s \\
Efficiency       & 0.4 \\
Traveltime       & 1.5 years \\
Acceleration when thrusting & 0.12 mm/s$^2$ \\
Fraction of thrusting arcs & 60\,\% \\
Power/mass of power system and thruster & 100 W/kg \\
Tank mass versus tank content (LOX) & 0.01 \\
\hline
$\rightarrow$ Propellant mass versus rock mass & 0.15 \\
$\rightarrow$ Dry mass versus rock mass & 0.043 \\
$\rightarrow$ Mass ratio & 23 \\
\hline
\end{tabular}
\label{tab:transferSC}
\end{table}

The assumed specific impulse of 2500 s is on the high end of Hall
thrusters and low end of gridded ion engines \citep{GoebelAndKatz2008}. The assumed efficiency
of 40\,\% is somewhat lower than the efficiency of state of the art
xenon Hall thrusters which are typically above 50\,\%. We motivate the
assumption by the fact that O$_2$ is less optimal propellant than
xenon.

The power per mass ratio of 100 W/kg is typical to contemporary solar
panel power systems. The thruster is typically quite lightweight in
comparison. We assume a passively cooled LOX tank. The tank walls can
be thin since the pressure is low and the tank does not have to
withstand launch vibrations or impulsive accelerations. We
obtain the result that 13\,\% of the initial rock must be turned into oxygen, so
that the ratio (propellant : rock) becomes (0.15 : 1). The 13\,\%
corresponds to only about one third of the total oxygen content
of typical asteroid rock.

Thus, the ratio of payload to dry mass of the transfer spacecraft is
23. If the spacecraft is used for multiple trips (1.5 year transfer
followed by shorter triptime for going back to the asteroid), the
effective mass ratio is increased.

\subsection{H$_2$O electric propulsion}

A drawback of the FFC Cambridge process is that the
CaCl$_2$ electrolyte must be imported from Earth.
Instead of extracting O$_2$ from rock,
one can use a (C-type) water-rich asteroid, extract water by heating,
and use the H$_2$O for Electric Propulsion. For Electric
Propulsion thrusters, O$_2$ and H$_2$O are rather similar.
Both are light molecules and when in hot and ionized state, they are chemically active.

The amount of water in the material is likely to be less than what is
needed for propulsion, unless the specific impulse is chosen to be
particularly high or the asteroid is particularly wet. Thus one must
probably be prepared for mining and drying more material than one
transports. To handle the waste material in the most sustainable way,
one can create an artificial asteroid of the abandoned material and
set it to orbit the parent asteroid. Then the dried-up material is not
wasted permanently, but it remains most easily accessible to those
future miners who prefer water-independent transportation techniques
such as FFC Cambridge.

Water is easier to store than O$_2$ because it is storable as a
room temperature liquid.




\subsection{Hydrogen reduction of oxides}

The Swedish-Finnish steelmaking company SSAB intends to replace traditional carbon
reducing agent in steelmaking by hydrogen, thus enabling CO$_2$-free
production of steel. Hydrogen gas reacts with iron oxides to form
reduced iron and water. The water is electrolysed to extract the
oxygen. The hydrogen is injected back into the reactor.

Ordinarily, hydrogen reduction is applicable only to iron oxides.
Turning the hydrogen to atomic or ionized form might also allow reduction
of other oxides \cite{SabatEtAl2014}. This is relevant for
asteroids because most of the oxygen is bound in silicon and magnesium
oxides.

The benefit relative to FFC Cambridge is that there is no need to
import process chemicals from Earth. Hydrogen is needed, but it can be
circulated, and the initial amount can be obtained from the water that
exists in most asteroids to some extent.

\subsection{Electric sail}

The solar wind electric sail (E-sail,
\citep{Janhunen2004,JanhunenEtAl2010}) is a propellantless propulsion method,
based on the momentum flux of the solar
wind. The E-sail consists of long and thin metallic tethers that
are kept in high positive potential by an onboard high voltage source
and electron gun that pumps out negative charge from the system. The maximum thrust of an
E-sail depends on materials and other parameters, but is roughly of the
order of $\sim 1 $ newton \citep{JanhunenEtAl2010}. With acceleration of 0.1 mm/s$^2$ as in
Table \ref{tab:transferSC}, the single transfer spacecraft can thus move 10
tonnes of material. To build a 200 person settlement weighing
78000 tonnes (Table \ref{tab:params200}), one needs $\sim 10^5$
trips. Thus one needs a large fleet of E-sail spacecraft.

The E-sails are tens of kilometers in diameter.  Thus, traffic
congestion at the asteroid and at the settlement construction site
become issues. The problem can be avoided by moving the materials
first by traditional spacecraft which rendezvous with E-sails once
there is enough free space around.  This increases the complexity to
some extent, but the scheme remains efficient since the majority of
the delta-v comes from the propellantless E-sail.

A fully autonomous optical navigation system is preferred. Otherwise
operations and radio communication costs can become high, because the
fleet is large. An autonomous optical navigation system was
in principle demonstrated already in Deep Space 1 in 1998.

The E-sail tethers are made of multiple wires to be tolerant of the
natural micrometeoroid flux. However, centimeter-sized particles can
break all the subwires of a tether at once. Debris \MARKII{possibly generated} by the
asteroid mining is thus potentially dangerous for the E-sails. Making
rendezvous with the E-sails far enough from the asteroid also mitigates
this problem.

\subsection{Comparison of methods}

Table \ref{tab:comparison} summarizes the benefits of the four
propulsion options for moving asteroid materials.

\begin{table}[htb]
\centering
\caption{Comparison of propulsion options for moving asteroid materials
  (Section \ref{sect:propulsion}).}
\begin{tabular}{lccccc}
               & O$_2$    & H$_2$O    & H$_2$   & E-sail \\
               & EP       & EP        & reduct. &        \\
\hline
Any asteroid   & +        &           & (+)     & +      \\
No waste       & +        &           &         & +      \\
No cryogenic tank &       & +         &         & +      \\
No Earth-imports &        & +         & +       & +      \\
Easily dockable& +        & +         & +       &        \\
No large fleet & +        & +         & +       &        \\
\hline
\end{tabular}
\label{tab:comparison}
\end{table}

\subsection{Space manufacturing to increase the mass ratio}
\label{subsect:advanced}

The Dutch-Luxembourgian company Maana Electric
(\url{http://www.maanaelectric.com}) is developing a self-contained
automatic factory, built in a standard-sized shipping container that
takes in desert sand robotically and produces finished solar panel
arrays, installing them in the surrounding desert. The company targets
not only Earth, but also the solar system. If this technology proves
to be practical, one could use it to produce solar panels for the
transfer vehicle from the mined asteroid regolith, thus reducing the
mass that must be brought from Earth. Structural parts of the transfer
vehicle may be possible to 3-D print from the metal-rich residue of
the FFC Cambridge process.

Making parts of the transfer vehicle from asteroid materials is
simpler than making parts of the settlement, because the transfer
vehicles are unmanned and redundant. Thus a failure of one of them is
not catastrophic. However, if strict quality checking standards are
imposed, then it becomes feasible to also make structural parts of the
settlement from asteroid materials. Our baseline structural material
is piano wire steel. Piano wire steel is 99\,\% iron, which is
abundant on asteroids. Other structural materials such as magnesium
alloys or basalt or silica fibers can also be considered.

\section{Costing of 200 person settlement}

Table \ref{tab:cost200} gives the masses and
launch costs of the 200-person settlement. We assume FFC Cambridge (subsection \ref{subsect:O2EP}) and
that the asteroid surface miners and the O$_2$ extraction factory
together can weigh up to 3000 tonnes. The contribution of surface miners is likely negligible in comparison with the factory. In subsection \ref{subsect:O2EP} we found
that at the current unoptimized prototype level of the FFC Cambridge
process, one unit of Earth-imported CaCl$_2$ electrolyte produces 2.7
units of O$_2$ per year. From Table \ref{tab:transferSC},
transportation of one mass unit of asteroid rock needs 0.15 mass units
of O$_2$ propellant. Thus, transportation of one mass unit of asteroid
rock needs $0.15/(5\times 2.7)$ = 0.011 units of Earth-imported CaCl$_2$, if the
production period is taken to be 5 years. Thus, transportation of 96576 tonnes of
rock needs 1062 tonnes of CaCl$_2$, which is 35\,\% of the 3000 tonnes
allowed for the O$_2$ factory. Recall that this is based on the
presently existing unoptimized FFC Cambridge process
of \citet{LomaxEtAl2019}.

\begin{table}[hbt]
\centering
\caption{Mass and launch cost budget for 200-person settlement, assuming O$_2$ propellant extraction but no other space manufacturing.}
\begin{tabular}{lrrl}
                 & Total   & Per pers.  & Source \\
\hline
Asteroid rock    & 86423 t &      432 t & Table \ref{tab:budget200} \\
\hline
Transfer s/c dry &  3716 t &       19 t & Table \ref{tab:transferSC} \\
Miner \& O$_2$ ex.& 3000 t &       15 t & See text \\
Steel            &  \MARKII{2270} t &       \MARKII{11.4} t & Table \ref{tab:budget200} \\
Other            &  \MARKII{493} t &      \MARKII{2.47} t & Table \ref{tab:budget200} \\
Earth to L5      &  \MARKII{9479} t &       \MARKII{47.4} t & Sum\\
Earth to LEO     & \MARKII{28437} t &      \MARKII{142} t & $3\times$ L5 \\
\hline
Launch/F9        & \$\MARKII{85}B   &    \$\MARKII{426}M  & \$3k/kg \\
Launch/Starship goal  & \$\MARKII{853}M  &    \$\MARKII{4.26}M  & \$30/kg \\
\hline
\end{tabular}
\label{tab:cost200}
\end{table}

In Table \ref{tab:cost200} we assumed -- conservatively -- that chemical propulsion is used
to push the payloads from LEO to L5 so that the mass originally
launched to LEO is 3 times larger than the mass that ends up in
L5. Falcon 9 costs \$3000/kg to LEO in the default partially reusable mode. The fully
reusable Starship rocket might be as much as 100 times more cost-effective (\$30/kg). Adopting that, the launch cost per
settler becomes \$4.2M.

Predicting the eventual per-kilogram cost of Starship or its
competitor is challenging at the moment. One recent
statement of SpaceX speculated with only \$2M cost per launch,
i.e.~\$13/kg \citep{TechCrunchNov2019}. To cover the uncertainty, in the Discussion part we
shall also consider an intermediate price case of \$300/kg.

If the launch is e.g.~20\,\% of the total cost (the other being
designing and building the settlement and the asteroid
mining chain for obtaining the radshield rock), each settler needs
an initial capital of $5\times4.2\mathrm{M} = \$21$M. Settlers can earn their
investment back and more, since once living at L5, they can
teleoperate the nearby robotic settlement production factory complex
in zero-delay mode. Thus the first group of settlers earns money by
producing new settlements. They do it more efficiently than from
ground because they avoid the 2.6 second free-space communication delay in
teleoperation.

As was remarked above in subsection \ref{subsect:advanced}, solar
panels and structural parts (including tanks) of the Electric
Propulsion transfer spacecraft could be made from asteroid
resources. Structural parts of the O$_2$ factory could also be be made
of asteroid-derived steel. Once proper quality control processes are
in place, structural parts of the settlement can also be made
of asteroid-derived piano wire steel. In Table \ref{tab:cost200B} we
show an example where space manufacturing has cut the net Earth-imported mass to 20\,\% of the original.

\begin{table}[hbt]
\centering
\caption{Same as Table \ref{tab:cost200} but with space manufacturing.}
\begin{tabular}{lrrl}
                   & Total  & Per pers.  & Source \\
\hline
Mining \& transfer & 6716 t &    33.6 t    & Table \ref{tab:cost200} \\
Steel              & \MARKII{2270} t &    \MARKII{11.4} t    & Table \ref{tab:cost200} \\
\hline
M\&T effective     & 1343 t &   6.7 t    & $0.2\,\times$ \\
Steel effective    &  \MARKII{454} t &   \MARKII{2.3} t    & $0.2\,\times$ \\
Other              &  \MARKII{493} t &   \MARKII{2.5} t    & Table \ref{tab:cost200} \\
Earth to L5        & \MARKII{2290} t &  \MARKII{11.5} t    & Sum \\
Earth to LEO       & \MARKII{6870} t &  \MARKII{34.4} t    & $3\times$ L5 \\
\hline
Launch/F9          & \$\MARKII{20.6}B  &  \$\MARKII{103}M     & \$3k/kg \\
Launch/Starship goal    & \$\MARKII{206}M &  \$1M      & \$30/kg \\
\hline
\end{tabular}
\label{tab:cost200B}
\end{table}

That is, under the stated assumptions space manufacturing reduces
launch costs by a factor of 4.2.

\section{Discussion and summary}

Radiation shielding dominates the mass of beyond-LEO settlements.  The
assumed 9 t/m$^2$ shield provides \MARKII{max} 20 mSv/year
\MARKII{equivalent dose} environment. If the shield is made thinner, the effective dose would
increase rapidly.

For large settlements where the shield thickness is negligible in
comparison with the sphere radius, the mass density of the radshield does
not matter. But for 200-person settlements it does play a
role. We assumed density of 2.6 g/cm$^3$ which is the same as the bulk
density of larger asteroids such as Eros. Reaching this density
probably requires some technical effort, for example making bricks out
of it by pressing, \MARKIII{sintering} or melting \MARKIII{\citep{Soilleux2019}}.

It is advantageous to separate the water and to put it in its own
layer, inward of the regolith. In this way, the hydrogen of the water
moderates the spallated neutrons so that they can be captured by a
thin layer of boron or other neutron absorber. A 2\,\% water content
is sufficient for moderating the neutrons. Water is also intrinsically
a better shield material than rock, so it is better if more is
available. To be conservative, in the calculations we assumed only
2\,\%.

For moving the materials from asteroids there are several propulsion
options.  Extracting O$_2$ from rock by FFC Cambridge and using it for
Electric Propulsion is a possible method.  A drawback is the necessity
to import the CaCl$_2$ electrolyte from Earth.  Finding a water-rich
asteroid and using H$_2$O as Electric Propulsion propellant is one of
the other alternatives.  The E-sail is also one option. It needs no
propellant extraction, but requires a large fleet size.

Tables \ref{tab:cost200} and \ref{tab:cost200B} show the launch cost
per person without and with space manufacturing, and with present
(Falcon 9) and future (Starship) launch vehicles. Introduction of space
manufacturing (under certain assumptions) reduces launch costs by
a factor of 4.2, and replacing Falcon 9 by the Starship cost goal of \$30/kg
reduces them by a factor of 100 (Table \ref{tab:costsummary}):

\begin{table}[htb]
\centering
\caption{Launch costs per settler with various launch prices and with or without space manufacturing.}
\begin{tabular}{lll}
LEO cost & Space & Per \\
& manufact.& person \\
\hline
\$3000/kg (Falcon 9) & No & \$\MARKII{426}M \\
\$3000/kg (Falcon 9) & Yes & \$\MARKII{103}M \\
\$300/kg  & No & \$\MARKII{42.6}M \\
\$300/kg  & Yes & \$\MARKII{10.3}M \\
\$30/kg (Starship goal) & No & \$\MARKII{4.26}M \\
\$30/kg (Starship goal) & Yes & \$1M \\
\hline
\end{tabular}
\label{tab:costsummary}
\end{table}

If Starship reaches \$30/kg, the first 200-person settlement could probably
be built without space manufacturing. Each settler might need initial
capital of $\sim \$21$M, 20\,\% of which is the launch. Such initial
capital sounds realistic for 200 settlers.  The money is used to build
and launch the parts of the settlement as well as the asteroid mining
infrastructure needed to make its radiation shields. Once the settlers
have moved in, they can earn back the money by selling new settlements
produced by their robotic mining and manufacturing infrastructure,
which they are now in the position to use more smoothly by delayless
teleoperation.

In Table \ref{tab:costsummary} we also list the LEO launch price
intermediate case of \$300/kg, which is 10 times smaller than Falcon
9, but 10 times higher than the Starship goal. Because Starship launches 150
tonnes per launch, \$300/kg corresponds to each Starship launch costing
\$45M, which is about the same as the present Falcon 9 launch cost in
its default partially reusable configuration (\$50M). Given that Starship
is fully reusable, its launch should cost less than that of Falcon 9,
because the fuel cost is only $\sim \$2$M per launch. In the \$300/kg
case, the launch cost per settler is \$\MARKII{10.3}M if space manufacturing is
used. Such figure sounds realistic for 200 settlers. If space
manufacturing is not used, then the launch cost is \$\MARKII{42.6}M per
settler. This initial cost level is also probably feasible for 200 settlers,
provided that there is a credible roadmap for lowering costs in the future by implementation of space
manufacturing and/or by lowering launch costs.

We summarize our main findings:
\begin{enumerate}
\item The radiation shield against GCRs dominates the mass (97\,\% in
  the case of 200 person settlement). It can be asteroid rock.
\item A sphere is the optimal shape for radiation shielding. Hence
  the two-sphere dumbbell configuration is best.
\item As structural material we recommend piano wire steel.
\item Several propulsion options to transport rock from asteroid to L5
  are viable.
\item To cut costs by space manufacturing, one can make solar panels and structural parts of the
  transfer vehicles and the settlement from asteroid materials.
\item The economic case looks promising if LEO launch cost is \$300/kg or below.
\item The settlers make money by constructing more settlements by
  teleoperating a nearby robotic factory with negligible communication
  delay.
\end{enumerate}

\section{Acknowledgement}

The results presented have been achieved under the framework of the
Finnish Centre of Excellence in Research of Sustainable Space (Academy
of Finland grant number 312356). I thank Thorsten
Denk for making the author aware of the problem that the lunar sling has due to
the Moon's rotation.





\begin{thebibliography}{00}

\bibitem[O'Neill(1974)]{ONeill1974}
O'Neill, G.K., The colonization of space, Physics Today 27, 9, 32--40
(1974).
\url{https://space.nss.org/the-colonization-of-space-gerard-k-o-neill-physics-today-1974/}

\bibitem[O'Neill(1977)]{ONeill1977}
O'Neill, G.K., The high frontier, New York (1977).

\bibitem[Globus et al.(2017)]{GlobusEtAl2017}
Globus, A., S.~Covey and D.~Faber, Space settlement: an easier way,
NSS Space Settlement J.~(2017). \url{https://space.nss.org/media/NSS-JOURNAL-Space-Settlement-An-Easier-Way.pdf}

\bibitem[Space News(2020)]{SpaceNewsAssureSpace}
``Assure Space won't cover collision risk in low Earth orbit'', Space
  News, March 11 (2020). \url{https://spacenews.com/assure-space-leaves-leo/}

\bibitem[Hou et al.(2015)]{HouEtAl2015}
Hou, X.Y., X. Xin, D.J. Scheeres and J. Wang, Stable motions around triangular libration points in the real Earth--Moon system,
Monthly Notices of the Royal Astronomical Society, 454, 4, 4172--4181 (2015). \url{https://academic.oup.com/mnras/article/454/4/4172/997896}

\bibitem[Globus and Strout(2017)]{GlobusAndStrout2017}
Globus, A., and J.~Strout, Orbital space settlement radiation
shielding, NSS Space Settlement J.~(2017). \url{https://space.nss.org/media/NSS-JOURNAL-Space-Settlement-Radiation-Shielding.pdf}

\bibitem[Kramer et al.(2004)]{FAX}
Kramer, R., H.J.~Khoury, J.W.~Vieira, E.C.M.~Loureiro, V.J.M.~Lima,
F.R.A.~Lima and G.~Hoff, All about FAX: A Female Adult voXel phantom for Monte
Carlo calculation in radiation protection dosimetry,
Phys.~Med.~Biol.~49, 23, 5203--5216 (2004).

\bibitem[Hassler et al.(2014)]{HasslerEtAl2014}
Hassler, D.M., et al., Mars' surface radiation environment measured
with the Mars Science Laboratory's Curiosity rover, Science, 343, 6169,
1244797 (2014).

\bibitem[Globus and Hall(2017)]{GlobusAndHall2017}
Globus, A.~and T.~Hall, Space settlement population rotation
tolerance, NSS Space Settlement J.~(2017). \url{https://space.nss.org/media/NSS-JOURNAL-Space-Settlement-Population-Rotation-Tolerance.pdf}

\bibitem[Yeomans et al.(2000)]{YeomansEtAl2000}
Yeomans, D.K., et al., Radio science results during the NEAR-Shoemaker
spacecraft rendezvous with Eros, Science, 289, 5487, 2085--2088 (2000).

\bibitem[O'Neill and Kolm(1980)]{ONeillAndKolm1980}
O'Neill, G.K.~and H.H.~Kolm, High-acceleration mass drivers, Acta
Astronaut., 7, 11, 1229--1238 (1980).

\bibitem[Baker and Zubrin(1990)]{BakerAndZubrin1990}
Baker, D.~and R.~Zubrin, Lunar and Mars mission architecture utilizing
tether-launched LLOX, AIAA 90--2109, AIAA/SAE/ASME/ASEE 26th Joing Propulsion
Conference, July 16--18, 1990, Orlando, Florida.

\bibitem[Landis(2005)]{Landis2005}
Landis, G.A., Analysis of a lunar sling launcher, J.~British
Interplanetary Soc., 58, 9/10, 294--297 (2005).

\bibitem[Eke et al.(2009)]{EkeEtAl2009}
Eke, V.R., L.F.A.~Teodoro and R.C.~Elphic, The spatial distribution of
polar hydrogen deposits on the Moon, Icarus, 200, 12--18 (2009).

\bibitem[Lomax et al.(2020)]{LomaxEtAl2020}
Lomax, B.A., M.~Conti, N.~Khan, N.S.~Bennett, A.Y.~Ganin and
M.D.~Symes, Proving the viability of an electrochemical process for
the simultaneous extraction of oxygen and production of metal alloys
from lunar regolith, Planet.~Space Sci., 180, 104748 (2020).

\bibitem[Andreussi et al.(2019)]{AndreussiEtAl2019}
Andreussi, T., E.~Ferrato, V.~Giannetti, A.~Piragino, G.~Cifali,
M.~Andrenucci and C.A.~Paissoni, Development status and way forward of
SITAEL's air-breathing electric propulsion, AIAA Propulsion and Energy
2019 Forum, 19--22 August 2019, Indianapolis, USA, AIAA 2019--3995,
doi:10.2514/6.2019-3995.

\bibitem[Lomax et al.(2019)]{LomaxEtAl2019}
Lomax, B.A., M.~Conti, N.~Khan, A.~Ganin and M.~Symes, The
Metalysis-FFC process for the efficient extraction of oxygen on the
lunar surface, Presentation given in ESA Space Resources Workshop, Luxembourg, October
10--11, 2019.

\bibitem[NASA website(2004)]{NASAwebsite}
\url{https://web.archive.org/web/20041024225737/http://neo.jpl.nasa.gov/risk/2000sg344.html}

\bibitem[Goebel and Katz(2008)]{GoebelAndKatz2008}
Goebel, D.M.~and I.~Katz, Fundamentals of electric propulsion: ion and
Hall thrusters, Wiley (2008). \url{https://descanso.jpl.nasa.gov/SciTechBook/st_series1_chapter.html}

\bibitem[Sabat et al.(2014)]{SabatEtAl2014}
Sabat, K.C., P.~Rajput, R.K.~Paramguru, B.~Bhoi and B.K.~Mishra,
Reduction of oxide minerals by hydrogen plasma: an overview, Plasma
Chem.~Plasma Process., 34, 1--23 (2014).

\bibitem[Janhunen(2004)]{Janhunen2004}
Janhunen, P., Electric sail for spacecraft propulsion, J. Prop. Power,
20, 763--764 (2004).

\bibitem[Janhunen et al.(2010)]{JanhunenEtAl2010}
Janhunen, P., P.K.~Toivanen, J.~Polkko, S.~Merikallio, P.~Salminen,
E.~H{\ae}ggstr\"om, H.~Sepp\"anen, R.~Kurppa, J.~Ukkonen, S.~Kiprich,
G.~Thornell, H.~Kratz, L.~Richter, O.~Kr\"omer, R.~Rosta, M.~Noorma,
J.~Envall, S.~L\"att, G.~Mengali, A.A.~Quarta, H.~Koivisto,
O.~Tarvainen, T.~Kalvas, J.~Kauppinen, A.~Nuottaj\"arvi and
A.~Obraztsov, Electric solar wind sail: Toward test missions,
Rev.~Sci.~Instrum., 81, 111301 (2010).

\bibitem[TechCrunch(2019)]{TechCrunchNov2019}
\url{https://techcrunch.com/2019/11/06/elon-musk-says-spacexs-starship-could-fly-for-as-little-as-2-million-per-launch/}

\bibitem[Soilleux(2019)]{Soilleux2019}
\MARKIII{Soilleux, R.J., Orbital civil engineering: waste silicates reformed into
radiation-shielded pressure hulls, J.~British Interplanetary Soc., 72, 244--252 (2019).}
















\end{thebibliography}



\end{document}